\documentstyle[12pt]{article}
\begin{document}
\pagestyle{plain}
\setcounter{page}{1}
\baselineskip16pt
\begin{titlepage}

\begin{flushright}
PUPT-1598\\
hep-th/9602135
\end{flushright}
\vspace{20 mm}

\begin{center}
{\huge Entropy and Temperature of Black 3-Branes}

\vspace{5mm}

\end{center}

\vspace{10 mm}

\begin{center}
{\large S.S.~Gubser\footnote{e-mail: {\tt ssgubser@puhep1.princeton.edu}}, 
I.R.~Klebanov\footnote{e-mail: {\tt klebanov@puhep1.princeton.edu}}
and A.W.~Peet\footnote{e-mail: {\tt peet@viper.princeton.edu}}}

\vspace{3mm}

Joseph Henry Laboratories\\
Princeton University\\
Princeton, New Jersey 08544

\end{center}

\vspace{2cm}

\begin{center}
{\large Abstract}
\end{center}

\noindent
We consider slightly non-extremal black 3-branes of type~IIB supergravity
and show that their Bekenstein-Hawking entropy agrees, up to a
mysterious factor, with 
an entropy derived by counting non-BPS excitations of the Dirichlet 3-brane.
These excitations are described in terms of the statistical
mechanics of a 3+1 dimensional gas of massless open string states.  This is
essentially the classic problem of blackbody radiation.  The blackbody
temperature is related to the temperature of the Hawking radiation.  We
also construct a solution of type~IIB supergravity describing a 3-brane
with a finite density of longitudinal momentum.  For extremal
momentum-carrying 3-branes the horizon area vanishes.  This is in agreement
with the fact that the BPS entropy of the momentum-carrying Dirichlet
3-branes is not an extensive quantity.

\vspace{2cm}
\begin{flushleft}
February 1996
\end{flushleft}
\end{titlepage}
\newpage
\renewcommand{\baselinestretch}{1.1} 

\renewcommand{\epsilon}{\varepsilon}
\def\equno#1{(\ref{#1})}
\def\equnos#1{(#1)}
\def\D#1#2{{\partial #1 \over \partial #2}}
\def\df#1#2{{\displaystyle{#1 \over #2}}}  
\section{Introduction}
\label{Intro}

Apart from their intrinsic importance, black holes\footnote{In this
short note we will not attempt to reference all of the developments in
the recent black hole literature.}  provide a testing ground for the
quantum theory of gravitation.  Classical General Relativity, together
with quantum field theory, implies that a black hole should be
assigned an entropy equal to one-fourth of its horizon area measured
in Planck units \cite{jb,swh}.  In a fundamental theory of quantum
gravity this Bekenstein-Hawking entropy should have a statistical
interpretation.  It has been argued \cite{oldbhss,ls,senbhss} that
string theory provides such an interpretation, because very massive
fundamental string states should form black holes, and the number of
such states exhibits the exponential Hagedorn growth.

Recently, a much improved understanding of the Ramond-Ramond charged
string solitons has emerged through the Dirichlet brane description
\cite{dlp,polchinski}.  This has led to rapid progress on the black
hole entropy problem.  In \cite{sv} a certain extremal 5-dimensional
black hole was constructed so that its horizon area is
non-vanishing. It was shown that the logarithm of its ground state
degeneracy, calculated with D-brane methods, precisely matches the
Bekenstein-Hawking entropy. This remarkable finding has been extended
in a number of directions. In \cite{bmpv} it was generalized to
rotating black holes. In \cite{cm} a similar 5-dimensional example was
considered, and it was further shown that the entropy of slightly
non-extremal black holes also matches the Bekenstein-Hawking
result. This allowed for a D-brane calculation of the temperature of
Hawking radiation. In \cite{hs1} similar results were obtained for
slightly non-extremal black strings in 6 dimensions (upon
compactification these strings reduce in a certain limit to the
5-dimensional black holes of \cite{sv}).

At this stage it is important to elucidate the criteria for agreement
between the D-brane and the Bekenstein-Hawking entropy, and to find
new successful examples. In this paper we provide a new and very
simple example of a black $p$-brane whose D-brane entropy almost matches
the Bekenstein-Hawking entropy.  This is the self-dual 3-brane in 10
dimensions.  Since it couples to the self-dual 5-form, it automatically
carries equal electric and magnetic charge densities.  A special property
of this object, as well as of those in \cite{sv}--\cite{hs1}, is that the
string coupling is independent of position.  Control over the value of the
string coupling at the horizon appears to be necessary for agreement
between the two definitions of entropy.  For $p$-branes with $p<3$ it is
easy to check that the D-brane entropy is not proportional to the horizon
area.  This is likely due to the string coupling becoming strong near the
$p$-brane.

The original 3-brane solution of type~IIB supergravity was constructed
in \cite{hs2}.  In section~2 we observe that at extremality this
solution has vanishing horizon area.  We construct a new class of
solutions describing 3-branes carrying finite momentum density along
one of its internal dimensions.  Although the longitudinal momentum is
known to stabilize the horizon area of extremal black strings
\cite{hs1}, here we find that it does not.  The fact that the
classical entropy is zero agrees with the fact that the logarithm of
the ground state degeneracy of the momentum-carrying Dirichlet
3-branes is not an extensive quantity.  In order to address objects
with non-vanishing horizon area, in section~3 we consider slightly
non-extremal 3-branes, whose masses satisfy $\delta M = M-M_0 \ll
M_0$.  To leading order in the parameter $\delta M/M_0$, which is a
measure of deviation from extremality, we find agreement between the
D-brane entropy and $1/4$ of the horizon area.  Amusingly, the
statistical mechanics of a non-extremal 3-brane is that of a photon
(and photino) gas in $3+1$ dimensions, which is the classic blackbody
radiation problem.  The scaling of entropy with energy may be derived
essentially from the well-known blackbody scaling laws,
\begin{equation}
M- M_0 \sim V T^4\ , \qquad\qquad S\sim V T^3\ .
\end{equation}
Working out the precise normalizations, we find that
the Bekenstein-Hawking and statistical entropies are not identical,
but are related by a mysterious proportionality factor. If, however,
only the transverse excitation modes of the 3-brane are counted,
then the statistical entropy becomes identical to the
Bekenstein-Hawking entropy. While this rule is suggestive, at the moment
we do not know how to justify it.

Upon coupling of the 3-brane to the 10-dimensional world, waves
colliding on the 3-brane may be converted to massless closed string
states.  This is Hawking radiation in the D-brane language \cite{cm}.
The blackbody temperature that one assigns to a non-extremal 3-brane
acquires the interpretation of the Hawking temperature.  In section~4
we conclude with a brief discussion.

\section{Entropy of 3-branes carrying longitudinal momentum}

The 3-brane solution to the equations of type~IIB supergravity was
originally obtained by Horowitz and Strominger \cite{hs2} and is given
by
\begin{eqnarray}
d s^2 &=& 
-\Delta_+ \Delta_-^{-1/2} dt^2 
+ \Delta_-^{1/2} (dx_1^2+dx_2^2+dx_3^2) 
+ \Delta_+^{-1}\Delta_-^{-1} dr^2 + r^2 d \Omega_5^2 \nonumber \\
F_{(5)} &=& Q (\epsilon_5 + *\epsilon_5) \nonumber\\
\Phi &=& {\rm const} \, .
\label{gensol}
\end{eqnarray}
In these equations $F_{(5)}$ is the Ramond-Ramond self-dual 5-form
field strength coupling to the 3-brane, and the dilaton field has an
arbitrary constant value for this solution.  We have also defined
\begin{equation}
\Delta_\pm(r) = \left (1-{r_\pm^4\over r^4}\right)  \, .
\end{equation}
The charge density on the 3-brane is 
\begin{equation}
Q = 2 r_+^2 r_-^2 \equiv 2 r_0^4 
\end{equation}
up to a convention-dependent proportionality constant.  In this  
section we will ignore such constants since they are irrelevant 
to our calculations.  For the
solution to be well-behaved, we need $r_+\ge r_-$. Extremality is
achieved when the horizon radius $r_+$ becomes equal to $r_-$.  The
extremal ADM mass is proportional to $Q$, as required by
supersymmetry.  The extremal solution preserves one-half of the ten
dimensional type~IIB supersymmetries, i.e. $N=1$.  We also introduce
an infrared cut-off by compactifying each internal coordinate $x^i$ on
a very large circle of radius $L$, i.e. imagine that the 3-brane is
wrapped around a large 3-torus $T^3$.

The 8-dimensional area of the horizon is 
\begin{equation}
A= \omega_5 r_+^5 L^3 \left[\Delta_-(r_+)\right]^{3/4} \ ,
\end{equation}
where $\omega_5=\pi^3$ is the area of a unit 5-sphere.  The classical
black 3-brane entropy,
\begin{equation}
S_{BH}={{A}\over{4}} \, ,
\end{equation}
therefore vanishes in the extremal limit.

If we fix the charge and consider a slightly non-extremal black
3-brane then, as we will see in the next section, the entropy of the
classical extremal black 3-brane scales as
\begin{equation}
S_{ext} \sim \omega_5 L^3 r_0^5 
\left[{{\delta M}\over{M_0}}\right]^{3/4} \, .
\end{equation}

In the case of the black string \cite{hs2}, which also had zero area
at extremality, it was possible to perform a boost along the string to
induce simultaneously finite ADM momentum and horizon area.

It is also easy to inject momentum $P$ along one\footnote{Note that
our conclusions would be unchanged if we performed additional boosts
involving any of the other spatial worldbrane directions.} of the
three spatial worldbrane directions, which we take to be $x^1$. The
appropriate solution may be found by performing a (now-standard) boost
on the solution \equno{gensol}.  In this way we obtain
\begin{eqnarray}
d s^2 &=& 
-\left(\cosh^2\alpha\Delta_+ \Delta_-^{-1/2} 
  - \sinh^2\alpha \Delta_-^{1/2} \right) dt^2 \nonumber\\
&+& \left(\cosh^2\alpha\Delta_-^{1/2} 
  - \sinh^2\alpha \Delta_+\Delta_-^{-1/2}\right) dx_1^2 \nonumber\\
&+& \sinh(2\alpha) \left(\Delta_-^{1/2} 
  - \Delta_+ \Delta_-^{-1/2}\right) dtdx_1 \nonumber\\
&+& \Delta_-^{1/2} (dx_2^2+dx_3^2)  
+ \Delta_+^{-1}\Delta_-^{-1} dr^2 + r^2 d \Omega_5^2  \, .
\label{boosol}
\end{eqnarray}
If we imagine that the $T^3$ is small, then we can think of the
configuration \equno{boosol} as a seven-dimensional black hole.  The
black hole has a gauge charge corresponding to the gauge field which
comes from the $(t,x^1)$ cross term in the metric.  Note that this
extremal solution is still BPS-saturated, as it preserves one
supersymmetry of a possible four (type~IIB compactified on $T^3$ to
$d=7$ has $N=4$ supersymmetry).  In ten dimensional language this
``charge'' is just the total ADM momentum, which is given by
\begin{eqnarray}
P_{ADM} &=& {{L^3\omega_5}\over{8\pi}} \sinh(2\alpha) (r_+^4-r_-^4)
\nonumber\\
&\equiv&{{2\pi n}\over{L}} \, , 
\end{eqnarray}
where $n$ is an integer and we are keeping the ten dimensional Newton
constant fixed.

If we let the deviation from extremality go to zero, but also take the
limit of infinite boost parameter, then for finite ADM momentum
\begin{equation}
P_{ADM} \sim L^3\omega_5 Q
\left[e^{2\alpha}{{\delta M}\over{M_0}}\right] \, ,
\end{equation}
we need the scaling ${\delta M}/M_0 \sim e^{-2\alpha}$.  

Then the entropy of a BPS-saturated state with this momentum number
$n$ is finite and given by
\begin{eqnarray}
S_{BPS} &\sim& 2\pi \sqrt{2n} \nonumber\\
&\sim& L^2\left[\omega_5 Q \right]^{1/2} 
\left[{{\delta M}\over{M_0}}e^{2\alpha}\right]^{1/2} \, .
\label{sbps}
\end{eqnarray}
This quantity is not extensive in the spatial worldvolume of the
3-brane.  The entropy density, measured per unit spatial worldvolume,
goes as
\begin{equation}
s_{BPS} \equiv {{S_{BPS}}\over{L^3}} \to 0 \, .
\end{equation}
For a Dirichlet $p$-brane, this zero BPS entropy will actually happen
for any value of $p>1$, as follows. A BPS-saturated excitation on the
worldvolume is effectively restricted to live in a single dimension,
because if there were two finite orthogonal momenta then the state
would no longer be BPS-saturated.  Therefore the scaling goes as
$S_{BPS} \sim \sqrt{n}$, while $P_{ADM}\sim L^{p}$, so that $n \sim
L^{p+1}$, and therefore
\begin{equation}
s_{BPS} \sim L^{(p+1)/(2p)}\to 0 \, .
\end{equation}
So we see that in order to have finite, nonzero ADM momentum and
finite, nonzero entropy, both measured per unit spatial worldvolume,
we need $p=1$, i.e. the string.

Let us now compare this conclusion about the Dirichlet 3-brane entropy
with results for the classical black 3-brane configuration.  Due to
the boost, we find that the Bekenstein-Hawking entropy of the
classical configuration \equno{gensol} is altered from its previous
value to
\begin{eqnarray}
S_{BH} &=& {{\omega_5}\over{4}} r_+^5 L^3 
\left[\Delta_-(r_+)\right]^{3/4} \cosh\alpha \nonumber\\
&\sim& \omega_5 L^3 r_0^5 
\left[{{\delta M}\over{M_0}}\right]^{3/4} e^\alpha
\end{eqnarray}
as $\alpha\to\infty$ and $\delta M/M_0\to 0$. Let us now take the
limit such that the ADM momentum remains finite.  Then we need the
scaling $\delta M/M_0 \sim e^{-2\alpha}$ and so the classical 3-brane
area goes as
\begin{equation}
A \sim e^{-3\alpha/2}e^\alpha \to 0 \, .
\end{equation}
This tells us that the BPS-saturated 3-brane with finite nonzero
momentum still has zero area.  Note that if we consider a modified
area given by the classical horizon area divided by
$\sqrt{g_{22}(r_+)g_{33}(r_+)}$, this scales similarly to the quantity
\equno{sbps}; however, it is difficult to give this modified area an
enlightening physical interpretation. 
\footnote{Note also that in the
above scaling limit $g_{tt}$ diverges on the horizon.  We thank
Gary Horowitz for pointing this out to us.}

Therefore we see that the entropy of the BPS-saturated classical
3-brane with momentum, which by definition is extensive in the horizon
area, is also zero. It is satisfying that the entropies on the
classical black 3-brane and Dirichlet 3-brane sides agree, as
expected.

\section{Statistical Mechanics of Non-extremal 3-branes }

In this section we will consider non-BPS excitations of the 3-brane.
In the D-brane picture the excitations we have in mind are described
by a dilute gas of massless open string states running along the brane
in arbitrary directions.  The average total momentum is zero.  The
momenta of the massless string states are quantized:
\begin{equation}
\vec{p} = \df{2\pi}{L} \vec{n}
\end{equation}
where $\vec{n} \in {\bf Z}^3$.  The mass of the excited 3-brane is 
\begin{equation}
M = M_0 + \delta M = \df{\sqrt{\pi}}{\kappa} L^3 + 
  \sum_{i=1}^k \df{2\pi}{L} |\vec{n_i}| + O(g) \ .     
\label{MassEx}
\end{equation}
Here $M_0$ is the mass of the extremal 3-brane \cite{polchnotes},
$k$ is the number of open strings, and 
\begin{equation}
\kappa = \sqrt{8 \pi G_N} = g \alpha'^2 \ .
\end{equation}
The $O(g)$ term in \equno{MassEx} accounts for interactions among the
strings.  The validity of counting these states and no others to
obtain the entropy of a non-extremal $p$-brane was discussed in
\cite{dm} for the case $p=1$, and the same arguments apply here.  In
particular, our ability to control the decay rate of the non-BPS
states by making $L$ large allows us to count these states reliably
with $g$ and hence $G_N$ finite.

Rather than calculating the degeneracy of excited 3-brane states at a
given $\delta M$ directly, let us instead consider the statistical
mechanics of massless open string states in the grand canonical ensemble.
The temperature $T$ will later be identified as the Hawking temperature,
but for now one can regard our ensemble calculations as a trick to figure
out the degeneracies of brane excitation levels.

For a system with $N$ massless boson and fermion physical degrees of
freedom, the correct partition function is
\begin{equation}
Z = \prod_{\vec{n} \in {\bf Z}^3} 
     \left( \df{1 + q^{|\vec{n}|}}{1 - q^{|\vec{n}|}} \right)^N   
\label{PartZ}
\end{equation}
where we have defined
\begin{equation}
q = e^{-2\pi/LT} \ .
\end{equation}
One expects $N=8$, but for now we leave it arbitrary.
The dynamics of these modes on the brane
is given by ${\cal N}=4$ supersymmetric pure Yang-Mills theory with gauge 
group $U(1)$ \cite{witten,arkady,green}.  
For our purposes, however, it is more
revealing to view this theory as ${\cal N}=1$ Yang-Mills plus six chiral
multiplets.  The chiral multiplets are associated with transverse
oscillations of the brane, while the gauge multiplet describes internal
degrees of freedom.  We will find that,
to obtain perfect agreement with the Bekenstein-Hawking
entropy, it is necessary to count only the modes of transverse
oscillation, hence setting $N=6$ in \equno{PartZ}.

What subtlety of the gauge dynamics might prevent the gauge degrees of
freedom from being enumerated along with the transverse oscillations?  
A.~Tseytlin has suggested to us the following interesting
mechanism \cite{Tseytlin}.  
If one imposes periodic boundary conditions on the gauginos
along the Euclidean time direction rather than the standard antiperiodic
boundary conditions, then the two physical gaugino degrees of freedom
introduce a factor $(1 - q^{|\vec{n}|})^2$ into the partition function,
exactly cancelling the gauge boson contribution, $(1 -
q^{|\vec{n}|})^{-2}$.  Thus the gauge dynamics becomes in effect
topological.  We look forward to exploring possible justifications and
consequences of this insightful guess for the gaugino boundary conditions.

Equation \equno{PartZ} includes $N$ bosonic and $N$ physical
fermionic modes, and in $3+1$ dimensions each fermion mode makes
$7/8$ the contribution of a boson mode to the entropy and energy (the
corresponding ratio in $1+1$ dimensions is $1/2$).  Using the
relations
\begin{equation}
\begin{array}{c}
F = -T \log Z  \\ 
E = T^2 \D{}{T} \log Z  \\
S = (E-F)/T 
\end{array}
\end{equation}
we find
\begin{eqnarray}
E = \df{\pi^2}{16}N L^3 T^4  \nonumber \\
S = \df{\pi^2}{12}N L^3 T^3 \ .                       
\label{CanonicalES}
\end{eqnarray}
At this point it is easy to see how things change when $n_w$ 3-branes
are stacked on top of one another.  The massless open strings can now
connect any two of the branes, so there are $n_w^2$ states for every
one state we had before.  In this context it is important to recall
that there is no binding energy among the 3-branes \cite{witten}, so
strings running between different branes really are massless.
Furthermore, when $L$ is large, it makes no difference whether we
consider $n_w$ singly wound branes or one brane wrapped $n_w$ times
around $T^3$: the asymptotic density of massless string states per
unit volume is unaffected by such changes in boundary conditions.

To recapitulate, the prescription for $n_w > 1$ is to consider $n_w^2$
(very weakly) coupled thermodynamic systems, each identical to the
$n_w = 1$ system treated above.  Thus \equno{CanonicalES} becomes
\begin{eqnarray}
E &=& \df{\pi^2}{16}N n_w^2 L^3 T^4  \nonumber \\
S &=& \df{\pi^2}{12}N n_w^2 L^3 T^3 \ .                       
\label{CNW}
\end{eqnarray}
The relation between $E$ and $S$ in the microcanonical ensemble is
determined by eliminating $T$ from \equno{CNW}:
\begin{eqnarray}
S =\df{2}{3} N^{1/4} \sqrt{\pi n_w} L^{3/4} E^{3/4} \, .
\label{Micro}
\end{eqnarray}
Setting $E = \delta M$ in \equno{Micro}, one obtains the entropy of 
non-extremal 3-branes with mass $M_0 + \delta M$.  Using the formula 
\cite{polchnotes}
\begin{equation}
M_0 = \df{\sqrt{\pi}}{\kappa} n_w L^3                       
\label{Mass0}
\end{equation}
one can show finally that 
\begin{equation}
S = \df{2}{3} N^{1/4} \pi^{7/8} n_w^{5/4} \kappa^{-3/4} L^3 
   \left( \delta M / M_0 \right)^{3/4} \ .                  
\label{FinalEnt}
\end{equation}
This expression for $S$ should be comparable to the Bekenstein-Hawking
entropy.  Let us therefore turn to the calculation of the horizon area
in the low-energy supergravity theory.

The ADM mass formula for the black $3$-brane described by the metric
\equno{gensol} is \cite{lu}
\begin{equation}
M_{ADM} = \df{\omega_5 L^3}{2 \kappa^2} (5 r_+^4 - r_-^4) \ . 
\label{MADM}
\end{equation}
Applying this formula to the extremal case $r_+ = r_- = r_0$ and
comparing with \equno{Mass0}, one finds
\begin{equation}
r_0^4 = \df{\sqrt{\pi}}{2 \omega_5} n_w \kappa  \ .         
\label{R0Calc}
\end{equation}
The RR charge remains unchanged as we perturb away from extremality,
so $r_- = r_0^2/r_+$.  Writing $r_+ = r_0 + \epsilon$, one finds from
\equno{MADM} that
\begin{equation}
\df{\delta M}{M_0} = 6 \df{\epsilon}{r_0} 
\end{equation}
to lowest order in $\epsilon$.  Thus the horizon area of the metric
\equno{gensol} is
\begin{eqnarray}
A &=& \omega_5 r_+^5 L^3 \left( 1 - \df{r_-^4}{r_+^4} \right)^{3/4} 
       \nonumber \\
  &=& 2^{9/4} \omega_5 r_0^5 L^3 \left( \df{\epsilon}{r_0} 
       \right)^{3/4} \nonumber \\
  &=& 2^{1/4} 3^{-3/4} \pi^{-1/8} (n_w \kappa)^{5/4} L^3 
       \left( \delta M/M_0 \right)^{3/4}
\end{eqnarray}
and the Bekenstein-Hawking entropy is
\begin{equation}
S_{BH} = \df{2 \pi A}{\kappa^2} 
  = 2^{5/4} 3^{-3/4} \pi^{7/8} n_w^{5/4} \kappa^{-3/4} L^3
       \left( \delta M/M_0 \right)^{3/4}\ .
\end{equation}

If we include all
eight bosonic and fermionic modes in the statistical mechanics
treatment of D-brane excitations, we obtain the following relation
between the entropies,
\cite{as}
\begin{equation}
S = \left( {\textstyle{{4}\over{3}}} \right)^{1/4} S_{BH} \ .
\end{equation}
While the scaling exponents agree perfectly, a mysterious numerical
factor appears. We do not understand why the statistical counting gets
so close, yet fails to reproduce the Bekenstein-Hawking entropy.
Note, however, that if we set $N=6$ then $S= S_{BH}$.
It is tempting to conjecture that a subtle modification of the world
volume dynamics, such as the twisted boundary conditions proposed by
Tseytlin, is responsible for this. The bottom line is that
an ideal gas on $6 n_w^2$ massless bosons and fermions on the world
volume reproduces the Bekenstein-Hawking entropy. The fact that
this number is $\sim n_w^2$ agrees with the enhanced symmetry of
coincident 3-branes. The necessary number is {\it smaller} than the
$8 n_w^2$ massless modes of the 
weakly coupled ${\cal N}=4$ $U(n)$ gauge theory. A resolution of this
puzzle may be related to the question of binding of the 3-branes.
If the $n_w$ parallel 3-branes form a marginal bound state, then
the number of massless modes is indeed reduced compared to
what is expected for unbound 3-branes. Although we do not know what
produces this bound state, we may speculate that it is related to
confinement.

A bonus we get for computing the entropy in the grand canonical
ensemble is that the blackbody temperature $T$ used in
\equnos{\ref{PartZ}-\ref{CNW}} is related to the Hawking temperature.
This is a trivial consequence of the relation $M = M_0 + E$ where $E$
is the energy of the gas of massless open strings.  We know from
ordinary statistical mechanics that $dE = T dS$ when $L$ is held
fixed.  But $dE = dM$, so the relation $dM = T_H dS_{BH}$ from black hole
thermodynamics leads immediately to
\begin{equation}
T_H = \left( \df{8}{3 \pi^2 n_w} \df{\delta M}{L^3} \right)^{1/4}  =
\left( {\textstyle{{N}\over{6}}} \right)^{1/4} T
\, .
\label{HawkingT}
\end{equation}
At first it seems surprising that the Hawking temperature should be
independent of the string coupling $g$.  But it becomes inevitable
when one realizes that $T_H \sim T$, since the properties of the dilute
gas of open string states characterizing the excitation of the D-brane
depend in no way on $g$.  The string coupling determines only the
degree of diluteness necessary to make our arguments valid.  It
remains a fascinating problem to derive this $g$-independent temperature
from a string perturbative calculation of the amplitudes for decay
processes of the excited 3-brane, similar to the scattering amplitudes
computed in \cite{us}.

\section{Discussion}

In this paper we have presented a very simple Dirichlet brane system whose
entropy is almost identical to the Bekenstein-Hawking entropy of the
corresponding low-energy supergravity solution.  This relation
is so miraculous that it clearly requires a deeper understanding.
How does classical type~IIB supergravity ``know'' the Planck formula for
blackbody spectrum?  Apparently it does.  The numerical factor relating the
statistical and Bekenstein-Hawking entropies poses a puzzle, however.  We
are inclined to regard this factor as a hint that we have yet to learn
everything about the dynamics of coincident 3-branes.  The suggestion
\cite{Tseytlin} to make the gaugino fields periodic in Euclidean time is a
simple way to obtain perfect agreement with the Bekenstein-Hawking formula,
but justification for this guess awaits a more thorough understanding of 
the worldvolume gauge field.

Motivated by \cite{cm} we would also like to show precisely how the 3-brane
blackbody temperature translates into the Hawking temperature of the
outgoing closed string radiation.  We hope to report on these issues in the
near future.

\section*{Acknowledgements}

We are grateful to C.G.~Callan, G.~Horowitz,
J.~Maldacena, R.~Myers, A.~Strominger and A.~Tseytlin 
for illuminating discussions.  The work of I.R.~Klebanov
was supported in part by DOE grant DE-FG02-91ER40671, the NSF
Presidential Young Investigator Award PHY-9157482, and the James S.{}
McDonnell Foundation grant No.{} 91-48.  The work of A.W.~Peet was
supported in part by NSF grant PHY-90-21984.  S.S.~Gubser was
supported by the Hertz Foundation.



\end{document}